\documentclass{astlb}
\usepackage{graphicx}
\usepackage{natbib}
\usepackage{color}

\newcommand{\lum}{erg s$^{-1}$}
\newcommand{\flux}{erg s$^{-1}$ cm$^{-2}$}

\usepackage{multirow}

\definecolor{darkblue}{rgb}{0,0,0.9}

\begin{document}

\journalinfo{2014}{40}{4}{177}[184]

\title{Relation between the X-ray and Optical Luminosities in Binary Systems
with Accreting Nonmagnetic White Dwarfs}

\author{M.~G.~Revnivtsev\email{revnivtsev@iki.rssi.ru}\address{1}, E.~V.~Filippova \address{2,1}, V.~F.~Suleimanov\address{3,4}
  \addresstext{1}{Space Research Institute of RAS (IKI), Moscow, Russia}
  \addresstext{2}{ISDC, University of Geneva, Switzerland}
  \addresstext{3}{Institut f\"ur Astronomie und Astrophysik, Universit\"at T\"ubingen, Germany}
  	\addresstext{4}{Kazan (Volga Region) Federal University, Kremlevskaya ul. 18, Kazan, Russia}
}

\shortauthor{M.G.Revnivtsev et al.}

\shorttitle{X-ray and Optical Luminosities of Dwarf Novae}

\submitted{November 26, 2013}

\begin{abstract}
  We investigate the relation between the optical (g-band) and X-ray (0.5–10 keV) luminosities of accreting nonmagnetic white dwarfs. According to the present-day counts of the populations of star systems in our Galaxy, these systems have the highest space density among the close binary systems with white dwarfs. We show that the dependence of the optical luminosity of accreting white dwarfs on their X-ray luminosity forms a fairly narrow one-parameter curve. The typical half-width of this curve does not exceed 0.2--0.3 dex in optical and X-ray luminosities, which is essentially consistent with the amplitude of the aperiodic flux variability for these objects. At X-ray luminosities $L_{\rm x}\sim 10^{32}$ \lum\ or lower, the optical g-band luminosity of the accretion flow is shown to be related to its X-ray luminosity by a factor $\sim$2--3. At even lower X-ray luminosities ($L_{\rm x}\sim 10^{30}$ \lum\  ), the contribution from the photosphere of the white
dwarf begins to dominate in the optical spectrum of the binary system and its optical brightness does not drop below $M_{\rm g}\sim 13-14$. Using the latter fact, we show that in current and planned X-ray sky surveys, the family of accreting nonmagnetic white dwarfs can be completely identified to the distance determined by the sensitivity of an optical sky survey in this region. For the Sloan Digital Sky Survey (SDSS) with a limiting sensitivity $m_{\rm g}\sim 22.5$, this distance is $\sim$400--600 pc.
  

\end{abstract}

\section{Introduction}
Close binary systems with white dwarfs (WDs)
constitute the largest family of star systems with
compact objects onto which matter is accreted. The
universally accepted name of such binary systems,
cataclysmic variables (CVs), owes its origin to significant
brightness variations during their outbursts,
which were the reason for isolating this class of objects
(see, e.g., \citealt{warner03}). The CV brightness
variations are currently believed to result from the
instability of mass transfer in accretion disks around
compact objects \citep{lasota01} This turns them into
very important laboratories to investigate the turbulent
and magnetic viscosities in space plasmas.

The interest in accreting WDs has only increased
in recent years owing to the fact that such binary
systems are among the possible progenitors of type Ia
supernovae \citep{hachisu96}, which, in turn,
are used for cosmological studies \cite[][]{riess98}. 
Various calculations show that the longterm
evolution of WD during the accretion of matter
can lead both to an increase in its mass and an explosion
as a supernova and to the ejection of part of
its envelope and long-term mass loss \citep{prialnik95,yaron05}. The time scales of
the long-term evolution of WDs are very large (see,
e.g., the review by \cite{howell01}). Therefore,
these questions can be answered only by studying the
whole populations of binary systems with WDs and
this means that a criterion for their efficient selection
among the various objects in the sky is needed.

CVs radiate in a wide energy range: from the
infrared and optical bands to the hard X-ray band.
The main emission components in the case of white
dwarfs with a weak magnetic field are: the companion
star (as a rule, a dwarf star with temperatures
below 3500--4000 K; see, e.g., the review by \citealt{knigge11}), the accretion disk radiating in the optical
and ultraviolet bands, the WD with an effective
temperature of at least 8000--10 000 К  \citep{townsley03,townsley09}, and the boundary layer near the WD surface radiating in
X-rays \cite[see e.g.][]{pringle79,patterson85}.

\begin{table*}[htb]

\caption{Parameters of the dwarf novae in the off state whose X-ray luminosities were determined here}
\medskip
\begin{center}
\begin{tabular}{lccccccc}
Source & Distance,&$m_g$&Flux& Flux&$\log L_{\rm x},$&Observatory&Dates\\
& pc& mag&0.5-2 keVВ&2-10 keV& \lum\ && \\
\hline
SW UMa&$164\pm22$&16.87&$5.6\pm0.9$&$5\pm2$&30.53&SWIFT&2011.12.18-19\\
&&&&&&&2012.08.26,28\\
T Leo &$101\pm13$&14.86&$65\pm2$&$94\pm3$&31.28&ASCA&1998.12.13\\
BZ UMa&$228\pm63$&16.37&$21\pm1$&$43\pm3$&31.52&SWIFT&2012.08.24-25\\
VW Hyi&$64\pm20$&13.9V&$32\pm2$&$42\pm 3$&30.55&ASCA&1993.11.08\\
WX Hyi&$260\pm64$&14.7V&$29\pm3$&$50\pm8$&31.80&Chandra&2002.07.28-29\\
SU UMa&$261\pm65$&14.6V&$20\pm2$&$39\pm5$&31.68&ASCA&1997.04.12-13\\
EF Tuc&$346\pm150$&14.5V&$20\pm1$&$67\pm5$&31.09&SWIFT&2008.09.12\\
RXJ1831&$980\pm630$&17.04&$5.7\pm0.7$&$14\pm2$&32.35&ASCA&1998.09.19-20\\
WW Cet&$158\pm43$&14.68&$68\pm4$&$125\pm3$&31.76&ASCA&1996.12.24-25\\
V405 Peg&$149\pm26$&16.79&$8.9\pm0.8$&$15\pm2$&30.80&SWIFT&2008-2012\\
TW Pic&$230\pm105$&15.2V&$40\pm1$&$149\pm7$&32.07&SWIFT&2007.11.13-12.31\\
\hline
\end{tabular}
\end{center}
\begin{list}{}
\item {\small
The optical g-band brightness of the binary systems was taken from the SDSS catalog \citep{ahn12}. The optical brightness
measurements for the binary systems marked by letter V are given in the V band (the V -band brightness of the optical systems in
the Vega system is very close to their g-band brightness in the AB system of the SDSS) and were taken from the catalogs either by \cite{bruch94} or by \cite{ritter}. The distances were taken from \cite{pretorius12}. The maximum
value in the case of asymmetric confidence intervals was taken as the uncertainty in the distance. The X-ray fluxes are given in units
of $10^{-13}$ \flux\ .
}
\end{list}
\end{table*}

If the rate of mass transfer through all parts of
the accretion flow (accretion disk, boundary layer)
is the same, then there must exist a certain relation
between their luminosities, which can facilitate the
binary system identification. However, the simple
picture of an accretion flow in the case of dwarf novae
in the off (quiescent) state is complicated by the fact
that the rates of mass transfer in different parts of the
accretion flow and, consequently, their luminosities
in different energy bands can be different \cite[see e.g.][]{wood86}. This conclusion is usually drawn
from the brightness temperature distribution along
the accretion disk radius in quiescence obtained by
eclipse mapping (see, e.g., \citealt{wood86,wood89}).
This distribution turns out to be considerably flatter
than is expected for optically thick disks with a
constant (along the radius) accretion rate \citep{ss73}. Nevertheless, as was shown,
for example, by \cite{biro00}, if the inner boundary
of an optically thick disk exceeds considerably the
WD size, then this conclusion may turn out to be
incorrect as a result of disk destruction by the WD
magnetic field, as in the case of intermediate polars,
or as a result of disk evaporation \citep{meyer94}. There are observational evidences
\cite{kuulkers11,revnivtsev12,balman12} that the inner disks of dwarf
novae in quiescence pass to the state of an optically
thin accretion flow with a plasma temperature of the
order of the virial one and radiate mainly in X rays.
Consequently, the relation between the rates of mass
accretion through an optically thick accretion disk
and onto the WD surface should be determined from
observations in all of the accessible spectral bands,
including the X-ray one.

Almost any CV detection strategy suffers from observational selection effects \cite[see e.g.][]{gansicke04,pretorius07,drake09}. The optical emission from a hot accretion disk ($T>6000-10000$ K) is difficult to distinguish in color from hot stars. Only the presence of strong Balmer emission lines in the spectra, along with the emission
in He II, CIII, and NIII lines (Bowen blend), can
serve as direct evidence. In optical sky surveys, the
systems are often identified as CVs if they exhibit
outbursts like dwarf novae or eclipses with a period
typical of CVs or the color is bluer than some color
in the short-wavelength part of the optical spectrum
\citep{green82} etc. (see, e.g., the strategy in
\citealt{szkody02}).

Among the optical sky surveys, the SDSS is
presently perhaps least affected by observational
selection effects; at the same time, record sensitivities
over large sky regions were achieved in it \citep{szkody11}. The disclosure of the long-predicted
population of CVs with short orbital periods has
already become the result of searching for CVs in the
SDSS \citep{gansicke09}.

In the currently existing X-ray sky survey, the
number of CVs is small \citep{verbunt97,schwope02,gansicke05,sazonov06,revnivtsev08,pretorius12,pretorius13} and is limited to the systems
with the highest X-ray luminosities/accretion rates,
as a rule, with magnetic WDs. Nevertheless, X-ray
sky surveys have a significant advantage in searching
for accreting WDs associated with their much larger
(compared to optical sky surveys) fraction in the
total population of detected sources and with cleaner
source selection criteria.

The population of accreting WDs is expected to be
dominant in the deep X-ray surveys of the Galactic
plane being conducted at present by the Chandra \citep{grindlay05,revnivtsev09,jonker11} and NuSTAR \citep{harrison13}
observatories and the expected sky survey of the
Spectrum-RG observatory \citep{pavlinsky09}.
Therefore, it is very important to determine the
algorithms that allow one to identify CVs maximally
reliably and completely among the enormous
number of sources in these new X-ray sky surveys.
In the existing X-ray sky surveys, the completeness
of CV identification among all of the detected
objects is actually achieved only after a complete
identification of all objects in the survey \cite[see e.g.][]{schwope02,sazonov06,revnivtsev08,pretorius12}.

The goal of this paper is to systematize our knowledge
of the relation between the X-ray and optical
luminosities of accreting nonmagnetic WDs for their
subsequent use in analyzing sky surveys. As a first
step in this direction, we restricted ourselves only
to CVs with nonmagnetic WDs. Such a selection
allows the additional complexities associated with the
contribution of cyclotron radiation to the optical and
infrared continuum of the sources to be avoided. We
investigate the systems in their typical states, i.e., in
those where they spend the vast bulk of their time and
will most likely be detected in random sky surveys.
For systems with a low accretion rate, dwarf novae,
we take the observations only in their quiescent/off
state.

\section*{Sample of sources}
To investigate the broadband emission from nonmagnetic
CVs, we combine three samples of sources:
dwarf novae from the ROSAT all-sky and North
Ecliptic Pole surveys \citep{pretorius12}, dwarf novae with measured parallaxes \citep{byckling10}, and low-luminosity nonmagnetic CVs
discovered in the SDSS \citep{reis13}.

\begin{figure}[htb]
\centerline{
\includegraphics[width=\columnwidth,bb=36 180 590 690,clip]{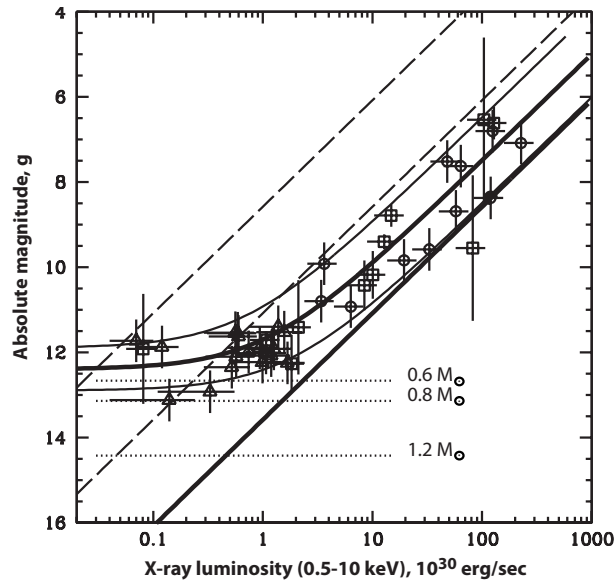}
}
\caption{Absolute magnitude of dwarf novae versus their X-ray luminosity. All measurements were taken during the off state (quiescence). The circles, squares, and triangles indicate the systems from the table, \cite{byckling10}, and \cite{reis13}, respectively. The thick straight line indicates the 1 : 1 optical/X-ray luminosity ratio ($\nu L_{\nu,g}/L_{\rm x} = 1$). The dashed straight lines indicate the 10 : 1 and 100 : 1 ratios. The horizontal dotted straight lines indicate the brightness of WDs
with masses of 0.6, 0.8, and 1.2 $M_{\odot}$ with temperatures of 8000--10 000 K. The thick solid curve indicates a simple analytical
fit; the thin solid curves indicate the region in which the X-ray and optical fluxes differ from the model by a factor of 1.6 and by
0.5 of the value, respectively.}
\label{dependence}
\end{figure}

Since \cite{pretorius12} provided only
the 0.5--2 keV fluxes for the sample of sources from
the ROSAT sky surveys, we performed an additional
analysis of the observational data from X-ray observatories
capable of covering the energy range 0.5--10 keV (ASCA, SWIFT, Chandra). The SWIFT
and ASCA data were analyzed with the LHEASOFT
software package. The Chandra data were processed
with the CIAO 4.5 software package. The energy
spectra of the sources were analyzed with the XSPEC
code \citep{xspec}. To fit the data, we used the
model of a multi-temperature optically thin plasma
$cevmkl$, which is commonly used to describe the
emission from such systems. The measured 0.5--2 keV
and 2--10 keV fluxes from the sources are given in the
Table 1.

\section*{Results}

\begin{figure*}[htb]
\begin{center}
\includegraphics[width=0.8\textwidth,bb=32 180 573 500,clip]{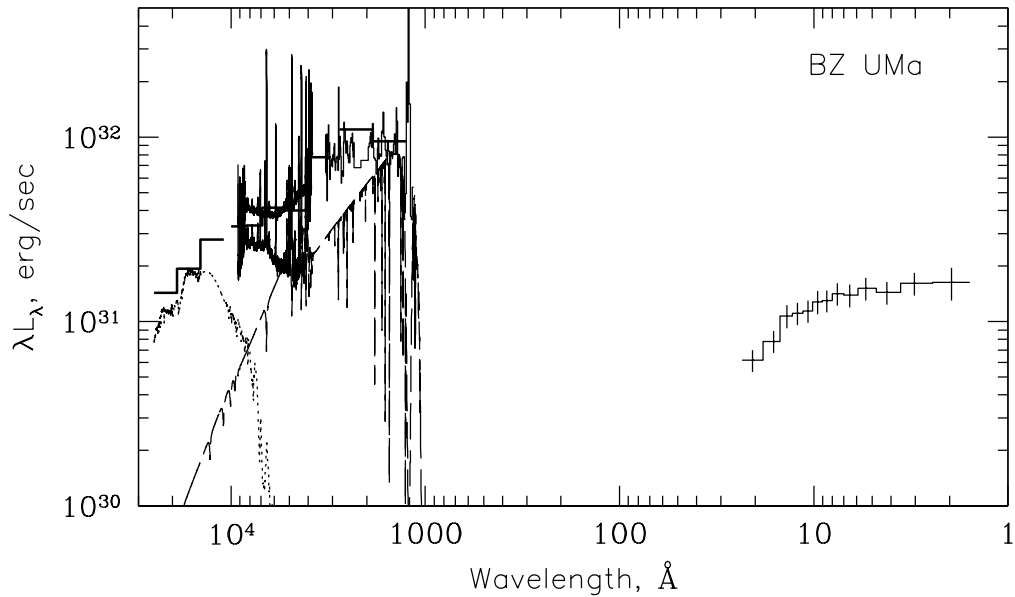}
\end{center}
\caption{Broadband spectrum of the dwarf nova BZ UMa in quiescence. The SDSS (optical range $\sim3800-9000$\AA), IUE
(ultraviolet range $\sim1100-3300$\AA , the measurements were grouped in $\sim$15\AA ), and SWIFT (X-ray range $\sim 1.5-20$\AA ) spectral
measurements are presented. The wide histograms indicate the photometric measurements from the 2MASS, SDSS, and
GALEX surveys. An approximate shape of the emission components associated with the contribution of the companion star
(a red dwarf with a temperature of 2950 K, the dotted curve) and the WD ($T = 15 kK$, $\log g=8$, the dashed curve) is shown
separately. The spectrum of the accretion flow obtained from the averaged spectrum by subtracting the contributions from the
white and red dwarfs is shown separately in the wavelength range 3700–9000 \AA . }
\label{bz_uma}
\end{figure*}

The correlations between the X-ray and optical
fluxes from CVs have been the subject of studies
starting from the systematic surveys of the Einstein
\cite[see e.g.][]{patterson85} and ROSAT \citep{vanteeseling96,verbunt97,schwope02} observatories. However, the
results obtained in recent years allow one to estimate
the distances to the systems being investigated more
reliably, to separate the systems in outbursts from
those in quiescence more reliably, and to study
fainter systems. New data have already revealed
various correlations between observed quantities (for
example, between the orbital periods of CVs and their
maximum brightness; \citealt{patterson11}).

Figure 1 presents the X-ray and optical luminosity
measurements for dwarf novae in quiescence from the
samples of sources being investigated. The X-ray
luminosities of the sources from \cite{byckling10}
were recalculated to the 0.5--10 keV energy band
using a coefficient $L_{\rm 0.01-100 keV}/L_{\rm 0.5-10 keV}\sim1.6$

When analyzing the dependence presented in
Fig. 1, we should take into account the fact that
the instantaneous optical and X-ray fluxes have an
uncertainty related to the stochastic flux variability
of the sources. The amplitude of this variability
was estimated from numerous X-ray observations
of the dwarf nova SS Cyg in quiescence (the X-ray
luminosity in this state is $L_{\rm x}\sim 2\times10^{32}$ \lum\ ) to
be about 30\%. It can be seen from Fig. 1 that the relation
between the X-ray luminosity of nonmagnetic
accreting WDs and their optical luminosity is fairly
close, actually unambiguous within the amplitude of
their chaotic variability.

The $L_{\rm opt}-L_{\rm x}$ relation in Fig. 1 has three main
peculiarities:

\begin{enumerate}
\item There are no dwarf novae in the off/quiescent
state with X-ray luminosities of more than $L_{\rm x}\sim
10^{32}$ \lum\ , while there are no obstacles to the
detection of such systems in various sky surveys.
Accreting WDs with high X-ray luminosities are actually
observed in various sky surveys, for example, in
the ROSAT \citep{pretorius13}, RXTE \citep{revnivtsev04,sazonov06}, INTEGRAL
\citep{revnivtsev08,krivonos12}, and SWIFT \citep{baugartner13} sky surveys, but
they all are magnetic WDs. The transition at high
mass accretion rates in the accretion disk to a state
with a different, much lower fraction of the X-ray
luminosity in the bolometric luminosity of the binary
system is probably responsible for the absence of
dwarf novae with high luminosities.
\item At X-ray luminosities $L_{\rm 0.5-10 keV}\sim
10^{31-32}$ \lum\ (at accretion rates onto the WD of
$\sim 10^{-12-11}$ $M_{\odot}$ yr$^{-1}$), the ratio $L_{\rm opt,g}/L_{\rm 0.5-10 keV}$ is
close to 2--3. The behavior of the optical/X-ray luminosity
ratio at $L_{\rm x}\sim 10^{31}$ \lum\ suggests that at this
or lower luminosities the accretion rate onto the WD
surface (which we observe as the X-ray luminosity of
CVs) is close to that in the outer accretion flow (which
we observe as the optical luminosity of CVs). A more
detailed study of the shape of the relation between the
optical luminosity of the accretion flow and its X-ray
luminosity requires a careful decomposition of the total
optical luminosity into its individual components, such as the contribution from the WD photosphere
and the contribution from the companion star, and
a proper allowance for the bolometric corrections for
various components.
\item At X-ray luminosities $L_{\rm x} <
10^{30}$ \lum\ , the optical luminosity ceases to depend
on the X-ray one. This peculiarity stems from the fact
that the contribution from the photosphere of the WD
itself, whose temperature never drops below 8000--
10 000 K \citep{townsley09}, begins to
dominate in the optical luminosity of CVs at low
accretion rates \cite[see e.g.][]{gansicke09,reis13}. Figure 1 presents the g-band absolute
magnitudes for WDs with masses of 0.8 $M_{\odot}$ and
0.6 $M_{\odot}$ whose photospheres have temperatures in
this range (taken from \citealt{holberg06,bergeron11}). It is important to note that
in CVs with X-ray luminosities below $10^{30}$ \lum\
the contribution of the accretion disk proper, which
occasionally can be separated from the total optical
flux by decomposing the the CV spectrum into
physically justified components \cite[see e.g.][]{littlefair08,reis13}, can be less than 10
the total optical flux and can lie on the extension of the
solid straight line in Fig. 1. However, confirming this
assumption requires a further study, which is beyond
the scope of this paper.
\end{enumerate}

{\sl Thus, the rate of mass accretion onto the
compact object in dwarf novae in quiescence
($\propto$ its X-ray luminosity) is comparable to the
rate of mass transfer to the outer accretion disk
($\propto$ the optical luminosity of the accretion flow).
} As a clear illustration of this fact, Fig. 2 presents a
broadband spectrum of the dwarf nova BZ UMa with
approximate designations of its individual emission
components. The spectrum of a red dwarf with a
temperature of 2950 K was taken from the library
of spectra by \citep{pickles98}. We calculated the
spectrum for a WD atmosphere of solar chemical
composition using our new computational code (for
more details, see \citealt{ibragimov03,suleimanov07,suleimanov13}).

\section*{Prediction for X-ray sky surveys}

The existence of a close relation between the
X-ray and optical luminosities of dwarf novae (in
quiescence) simplifies considerably the approaches
to identifying such X-ray sources being discovered
in new sky surveys. A simple analytical fit to
the observed data points in logarithmic coordinates
(without including the uncertainty in the quantities)
can be represented by the function

$$
M_g=-2.5\log\left(1.05\times10^{-5}+8.4\times10^{-6} L_{\rm x,30}\right),
$$

\begin{figure}[htb]
\includegraphics[width=1.1\columnwidth]{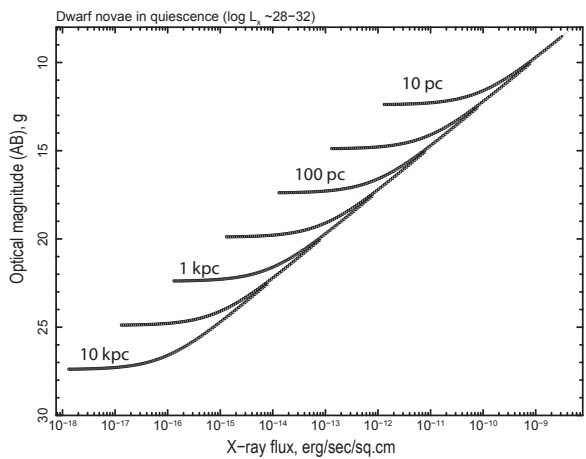}
\caption{Predictions for the ratio of the optical and X-ray fluxes from dwarf novae (in quiescence) in sky surveys. The objects
form a family of curves at various distances from the observer. The curves for some of the distances are shown.}
\label{fx_fopt}
\end{figure}

where $L_{\rm x,30}=L_{\rm x}/10^{30}$ \lum\ is the source’s X-ray
luminosity and $M_{\rm g}$ is its g-band absolute magnitude.
The predictions of X-ray and optical fluxes
for nonmagnetic dwarf novae using this dependence
for various distances are presented in Fig. 3. It
is clearly seen from the plots that the ratio of the
optical and X-ray fluxes from dwarf novae in a sky
survey can actually cover a range of approximately
two orders of magnitude (which is observed; see,
e.g., \citealt{schwope12}); such a spread arises from the
mixing of sources with different luminosities at different
distances and from the nonlinear dependence of
their optical luminosity on the X-ray one (note that
this situation differs greatly from the case of active
galactic nuclei; see, e.g., \citealt{sazonov12}). The
most important conclusion that we can draw from this
plot is that at a fixed sensitivity $m_{\rm g,lim}$ of the optical
survey accompanying the X-ray survey, there exists
some volume in which we can identify the complete
set of all dwarf novae detected in the X-ray band.
Indeed, since the limiting optical brightness of CVs at
an arbitrarily low accretion rate does not drop below
some value ($M_{\rm g}\sim 13$ for WDs with masses 0.6--
0.8 $M_{\odot}$), any, arbitrarily weakly accreting WDs will
be detected up to distances $\log d_{\rm pc}\sim 0.2(m_{\rm g,lim}-8)$.
For $m_{\rm g,lim}\sim20$, the boundary of the volume in which
we must assuredly detect any accreting nonmagnetic
WDs is about 250 pc. For a sensitivity $m_{\rm g,lim}\sim22.2$
corresponding to the SDSS, this boundary is pushed
to $\sim$690 pc. It should be noted, however, that this
limit depends significantly on the WD mass. For example,
at WD mass of 1.2$M_{\odot}$, its optical brightness
can be $M_{\rm g}\sim 14.4$ and the boundary of the viewed
volume is reduced to $\sim$360 pc. {\sl This fact should be
explicitly taken into account when analyzing the
completeness of the sample of accreting binary
systems.}

\section*{Conclusions}

We investigated the relation between the optical
and X-ray luminosities for a wide sample of accreting nonmagnetic WDs (dwarf novae) in quiescence. We
showed the following:

\begin{itemize}
\item The relation between the optical and X-ray
luminosities of dwarf novae (in quiescence)
forms a fairly narrow one-parameter curve.
The overwhelming majority of the investigated
system fit within 0.2 dex around the fitting
curve. At X-ray luminosities $L_{\rm 0.5-10 keV} \sim
10^{31}$ \lum\ (at accretion rates onto WD $\sim
10^{-1}$2 $M_\odot$ yr$^{-1}$), the ratio $L_{\rm opt,g}/L_{\rm 0.5-10 keV}$ is
close to 2−3.
\item The X-ray luminosity $(2-3)\times10^{30}$ \lum\ separates the regions in which the optical luminosity
of a binary system with an accreting
WD depends on its X-ray luminosity and the
region in which the optical luminosity of a binary
system is virtually independent of it. At
low X-ray luminosities, the contribution from
the WD photosphere dominates in the optical
luminosity of a binary system.
\item Based on this fact, we may conclude that
the family of accreting nonmagnetic WD can
be completely identified to the distances of
400--600 pc determined by the sensitivity
$m_{\rm g,lim}$ of the optical sky survey in this region
in current and future X-ray sky surveys.
\end{itemize}

\acknowledgements 
We used the data from the HEASARC electronic archive of the Goddard Space Flight Center (NASA). This work was supported by Program P21 of the Presidium of the Russian Academy of Sciences, Program OFN 16, the Dynasty Foundation, grant NSh-5603.2012.2, and the Russian Foundation for Basic Research (project nos. 12-02-01265, 13-02-00741, and 12-02-97006-r-povolzh’e-a).

{\sl Translated by V.Astakhov }

\end{document}